\documentclass[10pt]{bmc_article}

\usepackage{cite} 
\usepackage{url}  
\usepackage{ifthen}  
\usepackage{multicol}   
\usepackage[utf8]{inputenc} 

\def\includegraphics{}

\usepackage{graphicx}
\usepackage{graphics}

\newboolean{publ}

\newenvironment{bmcformat}{\begin{raggedright}\baselineskip20pt\sloppy\setboolean{publ}{false}}{\end{raggedright}\baselineskip20pt\sloppy}

\begin{document}
\begin{bmcformat}


\title{The Function of Communities in Protein Interaction Networks at Multiple Scales}
 

\author{Anna C F Lewis$^{1}$%
       \email{Anna C F Lewis - lewis@stats.ox.ac.uk}%
      \and
         Nick S Jones$^{2,3,4,5}$%
         \email{Nick S Jones - nick.jones@physics.ox.ac.uk}
      \and
         Mason A Porter$^{6,3}$%
         \email{Mason A Porter - porterm@maths.ox.ac.uk}
       and 
         Charlotte M Deane\correspondingauthor$^{1,5}$%
         \email{Charlotte M Deane - deane@stats.ox.ac.uk}%
      }


\address{%
    \iid(1)Department of Statistics, University of Oxford
    \iid(2)Department of Physics, University of Oxford
    \iid(3)CABDyN Complexity Centre, University of Oxford
    \iid(4)Department of Biochemistry, University of Oxford
    \iid(5)Oxford Centre for Integrative Systems Biology, University of Oxford
    \iid(6)Oxford Centre for Industrial and Applied Mathematics, Mathematical Institute, University of Oxford
}%

\maketitle


\begin{abstract}
        \paragraph*{Background:} If biology is modular then clusters, or communities, of proteins derived using only protein interaction network structure should define protein modules with similar biological roles. We investigate the link between biological modules and network communities in yeast and its relationship to the scale at which we probe the network.
      
        \paragraph*{Results:} Our results demonstrate that the functional homogeneity of communities depends on the scale selected, and that almost all proteins lie in a functionally homogeneous community at some scale. We judge functional homogeneity using a novel test and three independent characterizations of protein function, and find a high degree of overlap between these measures. We show that a high mean clustering coefficient of a community can be used to identify those that are functionally homogeneous. By tracing the community membership of a protein through multiple scales we demonstrate how our approach could be useful to biologists focusing on a particular protein.
        
	\paragraph*{Conclusions:} 
	We show that there is no one scale of interest in the community structure of the yeast protein interaction network, but we can identify the range of resolution parameters that yield the most functionally coherent communities, and predict which communities are most likely to be functionally homogeneous.


%
\end{abstract}

\ifthenelse{\boolean{publ}}{\begin{multicols}{2}}{}


\section{Background}

Large protein-protein interaction data sets \cite{shoemaker07a,Tarassov08, yu08} and functional information about many proteins are increasingly available. This allows one to investigate the patterns in protein-protein interactions that enable proteins to act concertedly to carry out their functions. In particular, considerable recent attention has been given to the modularity of the cell's functional organisation \cite{Hartwell, Ravasz, han2004edo}. A module is often thought of as a group of components that carry out a functional task fairly independently from the rest of the system. It is thought that such modules yield robust and adaptable systems \cite{alon2007isb}. There is also much suggestive evidence that modules within the cell are themselves the building blocks of a higher level of structural organisation (e.g. \cite{yook2004functional, hierarchicalmodularity1, hierarchicalmodularity2}). 

\mbox{}

Within the networks literature a great many algorithms have been proposed that locate dense regions in a network, often called communities (reviewed in \cite{comnotices, santo_review}). A community is loosely defined as a group of nodes that are more closely associated with themselves than with the rest of the network. Such communities are potentially good candidates for functional modules, and many studies report running one of the myriad algorithms for detecting community structure on protein interaction networks \cite{bu2003tsa,pereiraleal2004dfm, GOenrich, chen2006dfm, luo2007mop, mete2008saf, li2008gtm}. Having located communities, such studies then attempt to assess their functional homogeneity by searching for terms in a structured vocabulary ---usually the Gene Ontology (GO, \cite{GO}) or Munich Information Centre for Protein Sequences categories (MIPS, \cite{MIPS})---that are significantly over-represented within communities. If such terms exist, the identified communities are said to be `enriched' for biological function. In many studies such enriched communities are found, and hence are plausible candidates for biological modules.

\mbox{}

Recently there has been an acknowledgement that many community detection algorithms -- in particular all those that rely on optimising the quality function known as modularity -- impose an artificial \textit{resolution limit} on the communities detected \cite{reslimit}. Such algorithms return communities found at one particular resolution -- i.e. at one particular scale within the network -- whereas there are many scales of potential functional relevance within the protein interaction network. For example, one might expect to find smaller communities embedded inside progressively larger ones \cite{comnotices}. There are now algorithms available that include a `resolution parameter', which allow one to uncover structure at many different resolutions \cite{RandB, kumpula2007lra,heimo2008dmd,  arenas2008asc, blondel2008fuc}. However, no study to our knowledge has systematically applied such an algorithm and analysed the results across different resolutions in protein interaction networks (one study reports testing more than one value of a parameter akin to the resolution on a protein interaction network, in order to select an optimal value for their purposes \cite{pu2007identifying}).

\mbox{}

In this study, we probe the functional relevance of communities at multiple resolutions (scales) in the yeast protein interaction network, for two main biological reasons. First, considering the whole proteome, it is possible to view how the network breaks into communities (hierarchically or otherwise), and to investigate whether some scales of organisation are of more relevance than others biologically. Second, the relationship of multi-scale community structure to a particular protein is of interest: it is possible to see which other proteins co-occur with it at different resolutions -- perhaps it co-occurs robustly with a small group of proteins at high resolution but also with a larger set of proteins at a lower resolution. Both groups are of potential interest in understanding what role the protein plays. This is particularly pertinent for poorly annotated proteins, as their patterns of potential function can be revealed through clustering into communities \cite{y2h_worse}. 

\mbox{}

Although it is already thought that communities have some relationship to functional modules, here we expand on previous work to assess the functional relevance of communities in four main ways. 

\mbox{}

First, assessing functional relevance by counting over-represented terms amongst a group of proteins is not a sufficiently stringent test of functional relevance when the group of proteins in question is a community. This is because two proteins that interact are functionally more similar than a randomly chosen pair of proteins, so one must control for the number of interactions when assessing the biological relevance of a community (which will necessarily include more interacting pairs than a randomly selected group of proteins). We therefore control for the number of interacting proteins found in a community. 

\mbox{}

Second, instead of assessing functional homogeneity on a term by term basis we use all the annotations available within a given ontology.

\mbox{}

Third, GO and MIPS are subjective by their nature, both in the definition of the sets of terms themselves and in the process of annotation of terms to proteins. Due to their role in a particular process, a protein might well be both annotated more fully and have a higher probability of having had protein interaction experiments performed on it. Therefore, in addition to using GO and MIPS as protein functional characterizations, we use a single high-throughput experiment on the growth rates of gene knock-out strains under various conditions (using data from \cite{hillenmeyer2008chemical}).

\mbox{}

Fourth, protein interactions are of two fundamentally different types. The Molecular Interactions ontology \cite{MI} recognises two distinct types of interactions: physical associations (henceforth denoted $P$) and associations (henceforth denoted $A$). The main experimental type for the former are yeast-two-hybrid screens (e.g. \cite{li04}). The main type of experiment to fall under the latter are based on tandem affinity purification (TAP, e.g. \cite{collins08}). These interaction types are known to have very different properties \cite{shoemaker07a, vonmering02}. Additionally, the networks constructed using these two types of interactions have quite different global properties (see Table $1$). We thus investigate the two networks, based on type $A$ and type $P$ interactions, independently.

\mbox{}

We identify communities at multiple resolutions in these two fundamentally different interaction networks. We then use novel tests to determine the communities' functional homogeneity using three different characterisations of function. As the functional knowledge of proteins is far from complete (even for well characterised organisms such as yeast), we also search for topological properties of communities that are correlated with functional homogeneity.

\mbox{}

In our study we find many functionally homogeneous communities at multiple network resolutions. Almost all proteins are in functionally homogeneous communities at some resolution ($4652$ of $4980$ proteins in the $A$ network, and $5647$ of $5669$ proteins in the $P$ network). The resolution that places most proteins in functionally homogeneous communities is beyond the `resolution limit', or standard resolution, discussed above. At this maximum, $3071$ out of $4980$ proteins are in functionally homogeneous communities according to our GO similarity measure in the $A$ network. Communities at this resolution have mean size $73$, compared to mean size $293$ at the standard resolution. We find similar numbers for the $P$ network. Additionally, we find a high degree of overlap between communities judged functionally homogeneous using three separate quantifications of functional similarity. Through a further characterization of the communities using $26$ topological properties, we identify the mean clustering coefficient of a community as a good predictor of functional homogeneity, with a true positive rate of $70\%$ achievable with a false positive rate of $30\%$. In addition to these proteome-scale results, we demonstrate via examples how this approach can be used to predict groups of proteins likely involved in similar processes to a particular protein of interest. 

Additional Files can be found at \url{http://www.stats.ox.ac.uk/research/proteins/resources}.

\section*{Methods}

\subsection*{Protein-Protein Interaction Datasets}

Here we use the BioGrid (\url{www.thebiogrid.org}, downloaded January 2010, \cite{stark2006biogrid}), IntAct (\url{www.ebi.ac.uk/intact}, downloaded January 2010, \cite{IntAct}) and Mint databases (\url{mint.bio.uniroma2.it/mint}, downloaded January 2010, \cite{zanzoni02}) to assemble our protein interaction networks. We use only interactions between proteins that have an SGD identification (Saccharomyces Genome Database, \url{www.yeastgenome.org}). We divide interactions on the basis of their type ($A$ or $P$) and hence assemble the two networks (See Additional File $1$ for details). Of the potential $6607$ proteins in the yeast proteome (\url{www.yeastgenome.org}), there are $5002$ proteins connected by $A$ type interactions, and $5692$ connected by $P$ type interactions. Here we only study the largest connected component of these networks, leaving $4980$ proteins in the $A$ network and $5669$ in the $P$ network. Some summary statistics for the two amalgamated networks are shown in Table $1$. The $A$ network is denser, and has higher clustering.  There are $5947$ interactions in common between the $A$ and the $P$ networks.

\begin{table}
    \mbox{
	\begin{tabular}{|l|c|c|}
	\hline
	\textbf{Network} & \textbf{$A$} & \textbf{$P$}   \\ \hline
	Number of nodes & 4980 & 5669  \\
	Number of edges (of which self edges) & 48,330 (868) & 33,321 (941) \\ 
	Mean degree & 19.1 & 11.5  \\
	Mean clustering coefficient & 0.22 & 0.10  \\ \hline
	\end{tabular}
      }
\caption{\textbf{Network statistics of the $A$ and $P$ networks}}
\end{table}

\subsection*{Potts community detection}
We apply the Potts method \cite{RandB}. It partitions the proteins into communities at many different values of a resolution parameter, thus finding communities at different scales within the network. The method seeks a partition of nodes into communities that minimises a quality function (`energy'):
\begin{equation}
 H = -\sum_{ij} J_{ij}(\lambda) \delta(s_i, s_j),
\end{equation}
where $s_i$ is the community of node $i$, $\delta$ is the Kronecker delta, $\lambda$ is the resolution parameter, and the interaction matrix $J_{ij}(\lambda)$ gives an indication of how much more connected two nodes are than one would expect at random (i.e., in comparison to some null hypothesis). The energy $H$ is thus given by a sum of elements of $J$ for which the two nodes are in the same community. Optimising $H$ is known to be an NP-complete problem  \cite{NP, brandes2006mmh}, so one must use a computational heuristic. Here we use the greedy algorithm discussed in \cite{blondel2008fuc} and freely available (\url{www.lambiotte.be/codes.html}), which performs well against various benchmark tests \cite{santo09}.

\mbox{}

The interaction matrix $J$ has elements
\begin{equation}
J_{ij}(\lambda) = B_{ij} - \lambda R_{ij},
\end{equation} 
where the matrix $B$ with elements $B_{ij}$ is the adjacency matrix. In this case $B_{ij} = 1$ if proteins $i$ and $j$ interact, and $B_{ij} = 0$ otherwise. The matrix $R$ with elements $R_{ij}$ defines a null model, against which we are comparing the network of interest. Here we choose the standard {\it Newman-Girvan} null model \cite{newmanpre}, which has the property that it preserves the node degree sequence. That is, 
\begin{equation}
R_{ij} = \frac{k_i k_j}{2W},
\end{equation}
where $k_i =  \sum_{j} B_{ij}$ is the degree of node $i$, and $W = \sum_{ij} B_{ij} /2$ is the number of edges in the network. When $\lambda=1$, $H$ is the standard Newman-Girvan modularity quality function, upon which many community detection algorithms are based \cite{newmanpre, comnotices}. We hence refer to this value of the resolution parameter as the standard resolution. Values of $\lambda > 1$ probe the network at resolutions above the resolution limit.

\mbox{}

We investigate partitions of the network in the range $0.1 \leq \lambda \leq 1000$, and sample at intervals of $0.01$ on a logarithmic scale (we hence report results for $-1 \leq \log(\lambda) \leq 3$). At $\lambda = 0$, all nodes in our set will be assigned to the same community. As we increase $\lambda$, communities split and become smaller. If we allow $\lambda$ to increase until all of the entries in $J_{ij}$ are negative, then each node will be assigned to its own community. 

\subsection*{Pairwise measures of functional similarity}


It is impossible to uniquely quantify similarity in biological function. Here we rely primarily on the GO (\url{www.geneontology.org}), which provides the most comprehensive available database of functional annotations. We use the Biological Process sub-ontology annotations to yeast, which are maintained by the SGD consortium \cite{cherry1998sgd}. Terms are related to each other through a directed acyclic graph (DAG) (see Additional File $1$ Figure S$1$ for a visualisation of this structure). Proteins are annotated with the most specific terms that are known about them. It is then possible to add to this set their parent terms by following the structure of the DAG, up to the root node. Well-characterised proteins are those annotated with terms far from the root node. Of the $6346$ yeast proteins in the GO annotation set, $5347$ have biological process annotations (excluding the root node). We carried out the same tests using the Molecular Function and Cellular Component sub-ontologies, which gave similar results.

\mbox{}

We also use MIPS terms (\url{www.helmholtz-muenchen.de/en/ibis}, \cite{MIPS}), which are a useful double check on our results from GO, and have the added advantage that the terms are all found at the same level within the hierarchy of terms. Here we only use the top level of the MIPS hierarchy.

\mbox{}

Following \cite{pandey}, we quantify the functional similarity between two proteins $i$ and $j$ by finding the set of GO terms annotated to both proteins and counting the total number of proteins, $n_{ij}$, that share that set of terms. We then define a similarity measure between proteins $i$ and $j$ as 
\begin{equation}
 G_{ij}  = 1 -\log(n_{ij})/\log(N),
\label{eq:sim_measure}
\end{equation}
where $N$ is the total number of proteins. If both proteins are annotated with a set of terms that few proteins share, then they will be judged as functionally similar under this measure. Unlike many other measures, $G_{ij}$ does not penalise proteins for lack of annotation when judging their similarity. This is desirable, as we know that the GO annotations (even for the well-characterised {\it S. cerevisiae}) are far from complete. The quantity $M_{ij}$ is similarly defined through Equation \ref{eq:sim_measure} for the MIPS annotations. 

\mbox{}

The benefit of using a pairwise similarity measure that takes into account the full set of functional information available, rather than examining enrichment of function on a term by term basis, is that the measure has the potential to capture more general functional similarities between a pair of proteins.

\mbox{}

We also define a similarity between two proteins from a single high-throughput experiment via the growth rates of knock-out strains under a range of different conditions. Using the data in \cite{hillenmeyer2008chemical}, we define $C_{ij}$, the correlation in growth rates of the strain with gene $i$ knocked out to the strain with gene $j$ knocked out under $418$ different conditions:
\begin{equation}
C_{ij} = {\rm corr}(L_i, L_j),
\end{equation}
where the elements of the vector $L_i$ are
\begin{equation}
L_{i}^t = \log(\mu_{i}^c/\mu_{i}^t),
\end{equation}
the parameter $\mu_{i}^c$ is the mean growth rate of strain $i$ under different control conditions, and $\mu_{i}^t$ is the growth rate under one of the $418$ treatment conditions. We use the results from the homozygous strains. Because many gene deletions are lethal, there is only data available for $3625$ proteins, of which $3184$ are in the $A$ network and $3422$ are in the $P$ network.

\subsection*{Assessment of a community's functional homogeneity}
As mentioned previously, a fair test of the functional homogeneity of a community must take into account the fact that a pair of proteins that interact will be more similar than a randomly chosen pair. Standard enrichment tests do not take this into account, as they compare enrichment in a group of proteins, in this case a community, to what one would expect to attain from a randomly chosen set of proteins  \cite{boyle2004go}. A community necessarily contains many more interacting pairs than a randomly chosen set. We thus compare the pairwise functional similarities of all interacting pairs of proteins in a community to the same measure for all interacting pairs in the network, thereby controlling for the number of interacting pairs.

\mbox{}

To capture the pairwise similarity between two proteins that interact $\{ij\}$, we use $z$-scores:
\begin{equation}
z_{\{ij\}} = \frac{S_{\{ij\}} - \mu}{\sigma} 
\label{eq:zs}
\end{equation}
Where $S$ stands for one of our three similarity measures (based on GO, $G$, MIPS, $M$, or correlated growth rates, $C$), $\mu$ is the mean and $\sigma$ the standard deviation of all of the elements of $S$ for which proteins $i$ and $j$ interact in the network of interest ($A$ or $P$).

\mbox{}

A desirable quality for our test of functional homogeneity is the ability to compare communities found at different resolutions in an even handed manner. It is inherent in the nature of a statistical test that the significance of the test statistic under consideration (for example, the difference between the sample mean and the population mean) depends on the sample size: if one has a larger sample size, one can judge smaller differences to be `significant'. To determine the aggregate $z$-score, $z_{\rm agg}$, for the mean of a set of individual $z$-scores, $z_{\rm ind}$, one calculates $z_{\rm agg} = \sqrt N \mu(z_{\rm ind})$, where $N$ is the number of $z_{\rm ind}$s and $\mu(z_{\rm ind})$ is their mean \cite{stats_book}. So, given a $\mu(z_{\rm ind})$, a larger and hence more significant $z_{\rm agg}$ is achieved for a larger sample size (i.e. larger $N$). In order to separate out the effects of the number of interactors in the community from functional homogeneity, we thus choose to base assessment of functional homogeneity on the $\mu(z_{\rm ind})$, in our case $\mu(z_{\{ij\}})$ ($z_{\{ij\}}$ is defined in Equation \ref{eq:zs}). We judge as `significant' all those communities that have $\mu(z_{\{ij\}})$ above $0.3$, and call such communities ``functionally homogeneous''. We stress that this is not strictly an assessment of statistical significance, as we are choosing to ignore sample size. The value of $0.3$ would be judged to be significant at the $0.05$ significance level for any community with $30$ or more interacting pairs.

\subsection*{Topological properties that correlate well with functional similarity}
We investigate $26$ topological properties of the identified communities and assess whether any of these can be used to identify functionally homogeneous communities. Examples include mean clustering coefficient, betweenness measures, and network diameter. Any topological properties that correlate well with functional homogeneity can then be used to predict functionally homogeneous communities. We use each topological property as a classifier by predicting communities as functionally homogeneous when the value of that property is above a threshold, which we vary to construct a Receiver Operating Characteristic (ROC) curve. An ROC curve plots the number of communities correctly predicted as functionally homogeneous versus the number falsely predicted \cite{fawcett06}. We calculate the area under the ROC curve (AUC) for each metric at each value of $\lambda$, and report the mean of this quantity over resolutions between $0 \leq \log(\lambda) \leq 3$ (we exclude $-1 \leq \log(\lambda) < 0$, as the results are very noisy due to the small number of communities present). An AUC of $0.5$ would be expected from a random classifier.  AUCs of greater than $0.5$ imply that higher values of the metric are predictive of functional homogeneity. AUCs of less than $0.5$ imply predictive power if \textit{below} a threshold of that particular property was used (i.e. that the property and functional homogeneity are negatively correlated).

\section*{Results and Discussion}

\subsection*{Pairwise properties of proteins}

Community structure, if of any biological relevance, should uncover patterns that are more than the sum of effects from pairs of interacting proteins. In Table $2$ we show the pairwise similarity of proteins in each network under our three different measures of functional similarity (based on GO, MIPS, and correlated growth rates; see Methods). The similarity of pairs known to interact with either $A$ or $P$ type interactions is much higher than a randomly chosen pair of proteins under all three measures. This both helps motivate the investigation of the connection between functional similarity of proteins and the topology of the network, and demonstrates the necessity of taking into account pairwise properties when assessing any additional information that one can gain by studying communities.

\begin{table}
    \mbox{
	\begin{tabular}{|c|c|c|c|c|} 
\hline
	&\multicolumn{2}{c|}{\textbf{$A$}}&\multicolumn{2}{c|}{\textbf{$P$}}\\
	  & All pairs & Interacting pairs & All pairs & Interacting pairs  \\ \hline
	$G$ & 0.04 & 0.14 & 0.04 & 0.12  \\
	$C$ & 0.19 & 0.35 & 0.18 & 0.33  \\
	$M$ & 0.22 & 0.28 & 0.22 & 0.27   \\ \hline
	\end{tabular}
      }
\caption{\textbf{Pairwise similarities of proteins in the $A$ and $P$ networks under the three different similarity measures, $G$, $C$, and $M$}}
\end{table}

\subsection*{Communities}

\begin{figure}
\includegraphics[width=120mm]{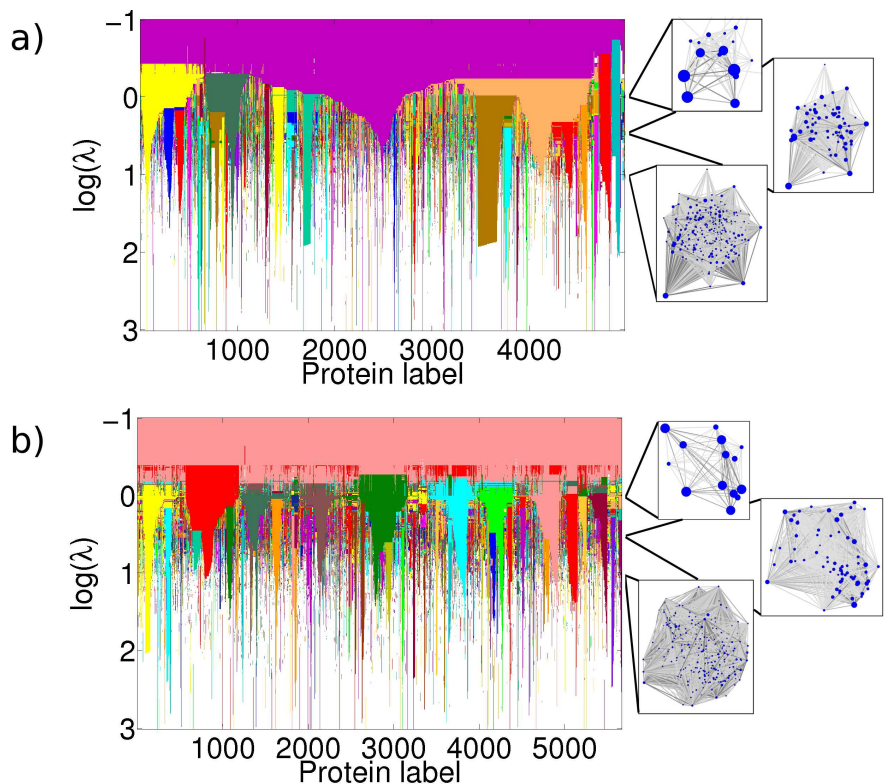}
\caption{\textbf{Communities identified in the $A$ and $P$ Networks}. Communities identified in the yeast protein interaction network for interactions of a) type $A$ and b) type $P$. When the resolution parameter $\lambda$ is very small,  all nodes are assigned to the same community (which is analogous to viewing the network at a great distance). As $\lambda$ is increased (viewing the network at progressively closer distances), more structure is revealed. The figures on the right hand side show visualisations of the networks' partition into communities at three different values of $\lambda$. Each circle represents a community, with size proportional to the number of proteins in that community, positioned at the mean position of its constituent nodes. (These positions were determined via a standard force directed network layout algorithm \cite{kamada}.) The shade of the connecting lines is proportional to the number of links between two communities. The main figure shows the communities that we find as we vary the resolution. We identify communities as the same through changing resolution parameter, and hence colour them the same, according to a convention described in Additional File $1$ (only communities of size $50$ or more are shown). Note that the ordering of proteins is not the same in the two figures. }
\end{figure}

Figure $1$ shows the communities that we find in the $A$ and $P$ yeast networks as the resolution parameter $\lambda$ is varied. As $\lambda$ increases, more and smaller communities are found (see Table $3$).  At $\lambda = 1$ (i.e. $\log(\lambda) = 0$), which corresponds to standard Newman-Girvan modularity \cite{newmanpre}, most communities contain a few hundred proteins. By $\log(\lambda) = 3$ however, almost all proteins are in communities of size three or smaller. As shown in Figure $1$, some sets of nodes are classified in the same community through large changes in the resolution parameter and hence represent particularly inter-connected parts of the network. Figure $1$ should be contrasted with Figures S$2$ in Additional File $1$, which are similar calculations on a random network and a network designed to possess strong communities. In the former, not much structure is present, in the latter, there are very distinct blocks.

\begin{table}
    \mbox{
	\begin{tabular}{|l|p{2cm}|p{2cm}|}
	\hline	
\textbf{$\log(\lambda)$}&\multicolumn{2}{c|}{\textbf{mean size of communities}}\\
 & \textbf{$A$} & \textbf{$P$}   \\ \hline
	-0.5 & 681 & 2834  \\
	0 & 293 & 405 \\ 
	0.5 & 73 & 79 \\  
	1 & 22 &  26 \\
	1.5 & 11 & 10  \\
	2 & 6 & 6  \\
	2.5 & 5 & 5  \\
	3 & 4 &  4 \\ \hline
	\end{tabular}
      }
\caption{\textbf{Mean size of communities in the $A$ and $P$ networks}}
\end{table}

\mbox{}

Figure $2$a illustrates for the $A$ network the number of communities of size four or more as the resolution changes, and Figure $2$b shows how many proteins are in those communities. (Figure S$3$ in Additional File $1$ is the same plot for the $P$ network, and shows similar behaviour). 

\mbox{}

The two networks, $A$ and $P$, contain very different types of interactions, and they can therefore be used to identify different aspects of the cell's functional organisation. The $A$ network is also much denser than the $P$ network. $A$ interactions would therefore dominate the clustering into communities, thereby making it very hard to pick out any structures given by $P$ type interactions (as occurs in \cite{morethanmodules}). When considering a particular protein or set of proteins, comparisons between communities found in the $A$ and $P$ networks can be made, see the Examples section. Global comparisons between the partitions of the $A$ and $P$ networks at a particular resolution are not necessarily meaningful as, for example, the size of communities depends both on the size and other properties of the network.

\mbox{}

Data files containing the $A$ and $P$ networks and the community membership of proteins at multiple resolutions are available at \url{http://www.stats.ox.ac.uk/research/proteins/resources}.

\begin{figure}
\includegraphics[width=120mm]{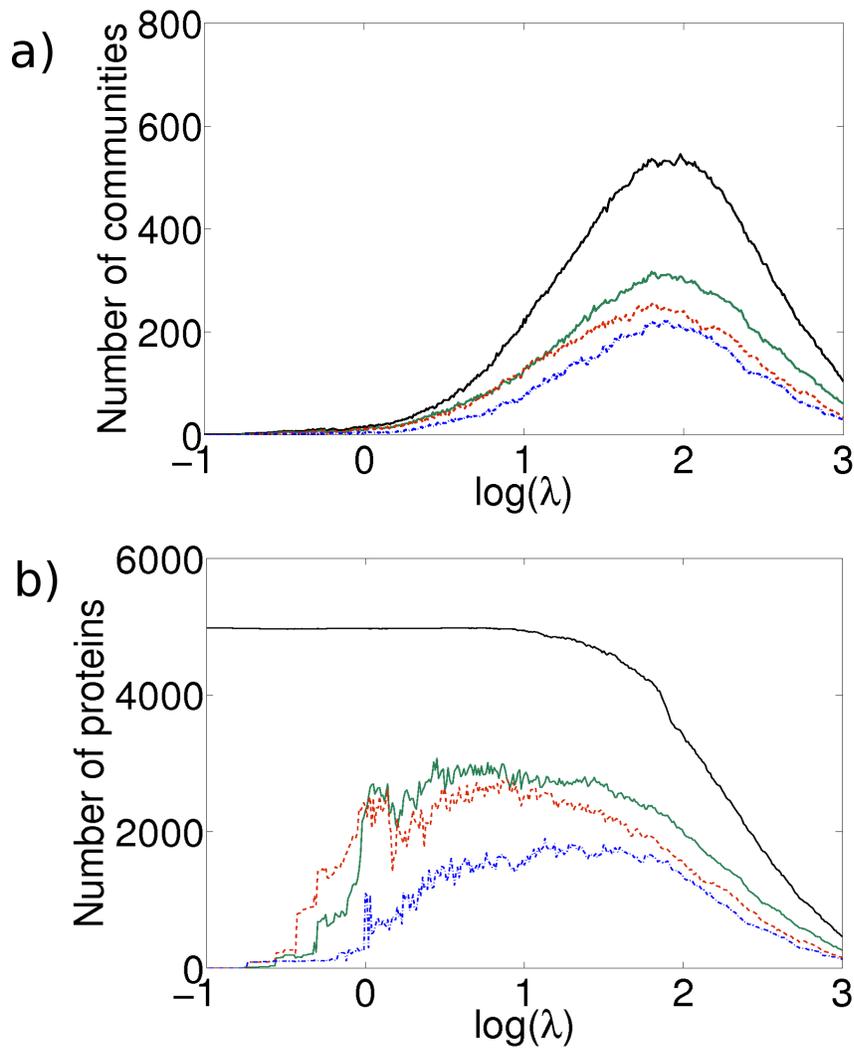}
\caption{\textbf{For the $A$ network a) the number of communities of size four or more and b) the number of proteins in such communities and the fraction of these that are judged functionally homogeneous.} a) The number of communities with changing resolution parameter (solid black curve) b) The number of proteins $p$ in communities of size four or more (solid black curve). Also shown are the numbers of communities/proteins in such communities judged to be functionally homogeneous according to the GO similarity measure (green curves), the MIPS measure (dot-dashed blue curves) and the correlated growth similarity measure (dashed red curves). At values of $\log(\lambda) \leq 0.5$, relatively few proteins are in communities judged to be functionally homogeneous. The equivalent figure for the $P$ network is given in Additional File $1$ Figure S$3$.}
\end{figure}

\subsection*{Functional homogeneity of communities}

We now assess how many communities are judged functionally homogeneous, looking in particular at how our results vary with resolution parameter.

\mbox{}

Figure $2$a illustrates the number of communities judged to be functionally homogeneous, and Figure $2$b shows the number of proteins in communities judged to be functionally homogeneous. Both are for the $A$ network. We find that the large communities present at small values of the resolution parameter $\lambda$ are not judged to be functionally homogeneous. As $\lambda$ is increased, larger numbers of proteins occur in functionally homogeneous communities, peaking in the range $1.5 < \log(\lambda) < 2$. At $\log(\lambda) = 1.5$, the mean community size is $73$ proteins, and the majority of proteins, $3071$ of $4980$, are in functionally homogeneous communities as judged by our GO similarity measure. The shapes of the curves of both Figure $2$a and b for all three similarity measures are very similar. Indeed, we find that the overlap between the communities judged to be functionally homogeneous between any two of the three measures is high (see Figure S$4$ in Additional File $1$); for example, it is $70\%$ between the GO and correlated growth rates measure over almost the entire range of the resolution parameter in both $A$ and $P$ networks. Given that the correlated growth similarity measure represents a very different data type to the GO and MIPS annotations, this agreement gives us confidence in the similarity measure we use for GO and MIPS. As we use only the top level of the MIPS functional annotations, we capture less information than the GO measure, so it is unsurprising that fewer communities are found to be functionally homogeneous under this measure.


\mbox{}

The $P$ network (see Figure S$3$ in Additional File $1$) shows a similar pattern to the $A$ network. One difference is that communities start to be judged as functionally similar at a slightly lower resolution. This is most likely due to the different topological properties of the $P$ network. That there are comparably many functionally homogeneous communities in the $P$ network as the $A$ network is of interest, as communities found in $P$ networks are found to be poor choices for predicting function on the basis of enrichment of terms \cite{y2h_worse}. 

\mbox{}

For almost all proteins, there is some value of the resolution parameter that assigns them to a functionally homogeneous community. In fact $4652$ out of $4980$ $A$ proteins and $5647$ and of $5669$ $P$ proteins are in such communities at some value of the resolution parameter. For a given protein, it may not be that it interacts most closely with proteins involved in the same process. Indeed it is often necessary to look at a larger scale, placing the community in a bigger community in order to identify the biological processes it participates in. Whether or not this is the case, and which network scale (resolution) is most indicative of the processes a protein is involved in, will depend on the particular protein one is interested in. This demonstrates the biological motivation for investigating community structure at multiple resolutions, and suggests the desirability of a method to easily identify those communities most likely to be functionally homogeneous.

\mbox{}

We might expect proteins involved in particular processes to show different propensities to lie in functionally homogeneous communities. We focus on a small but broad set of protein types, which are the GO biological process terms within the yeast GO slim \cite{GO_slim} that are annotated to at least $200$ yeast proteins. There are $11$ such terms, which are listed in Additional File $1$, as well as the numbers of proteins annotated to each. We investigate what fraction of each type of protein lie in communities judged functionally homogeneous under the GO measure through changing resolution parameter. Figure $3$ illustrates for the $A$ network these percentages for four particular processes. (Figure S$5$ in Additional File $1$ shows the same figure for all $11$ terms for the $A$ network and separately for the $P$ network). Proteins of some types are far more likely to be found in functionally homogeneous communities than others. For example, for both the $A$ and $P$ networks, proteins involved in chromosome organisation are far more likely to be found in functionally homogeneous communities than proteins involved in lipid metabolism. In addition, there are some indications that the resolutions of most interest can depend on the type of protein under investigation. As can be seen in Figure $3$, proteins involved in RNA metabolic processes are more likely to be found in functionally homogeneous communities at $\log(\lambda) = 0.8$, where the mean size of communities is $30$. In contrast, proteins involved in vesicle-mediated transport are found in greater numbers in functionally homogeneous communities at $\log(\lambda) = 1.7$, where the mean size of communities is $10$. 

\begin{figure}
\includegraphics[width=120mm]{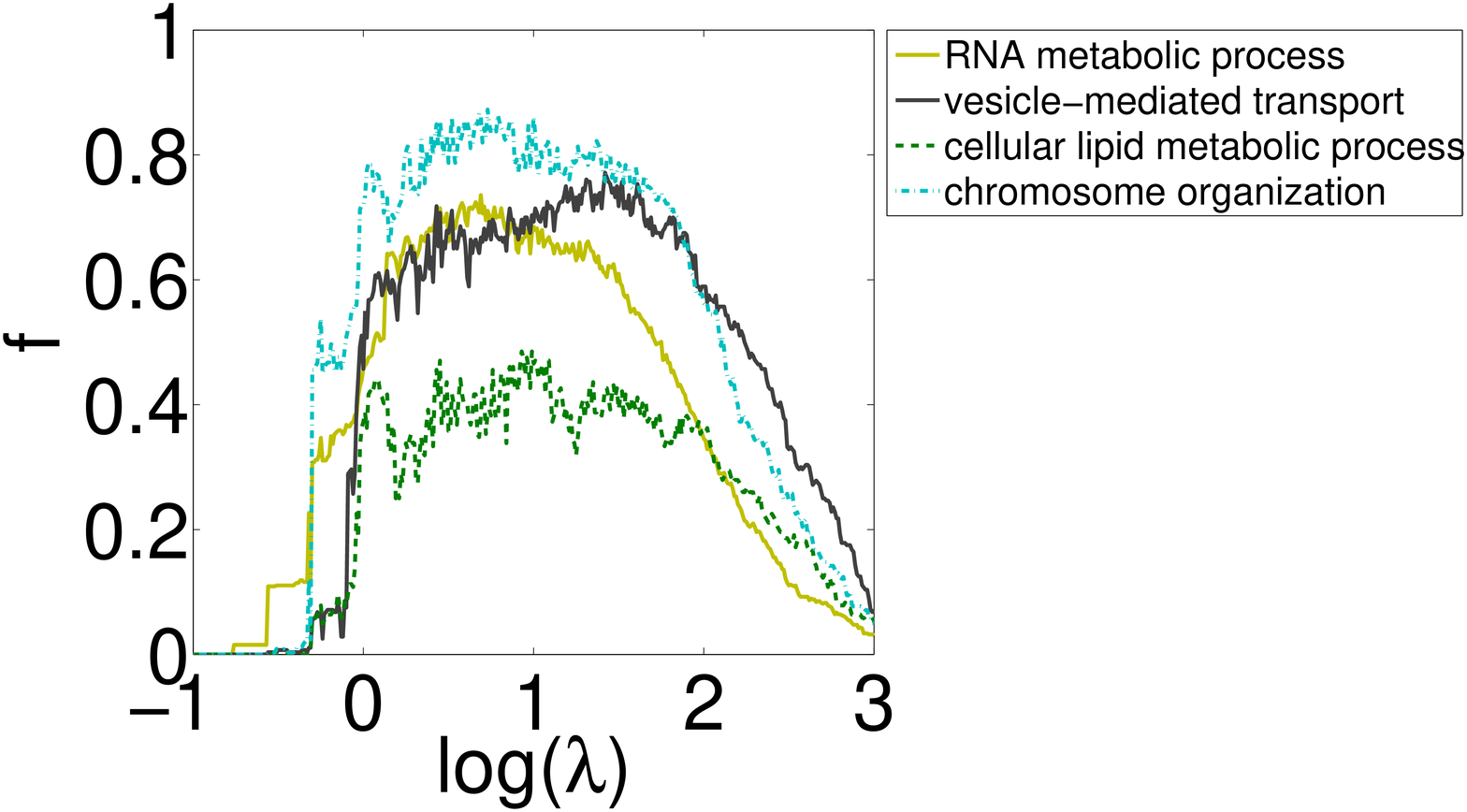}
\caption{\textbf{Fraction of proteins of particular types in functionally homogeneous communities.} The fraction of proteins, $f$, of particular types that are in functionally homogeneous communities in the $A$ network, with changing resolution parameter. With changing resolution parameter proteins of particular types have consistent differences as to how often they are found in functionally homogeneous communities. For example, proteins involved in chromosome organisation are far more likely to be in functionally homogeneous communities than proteins involved in metabolism. There are also some features that suggest `good' resolutions for particular processes. For example, a good resolution for proteins involved in vesicular mediated transport would be $\log(\lambda) = 2.7$ (for which the mean size of communities is $10$), whereas for proteins involved in RNA metabolic processes, $\log(\lambda) = 0.8$ would be better (the mean size of communities is $30$).}
\end{figure}

\subsection*{Examples of communities found at multiple resolutions}

\begin{figure}
\includegraphics[width=120mm]{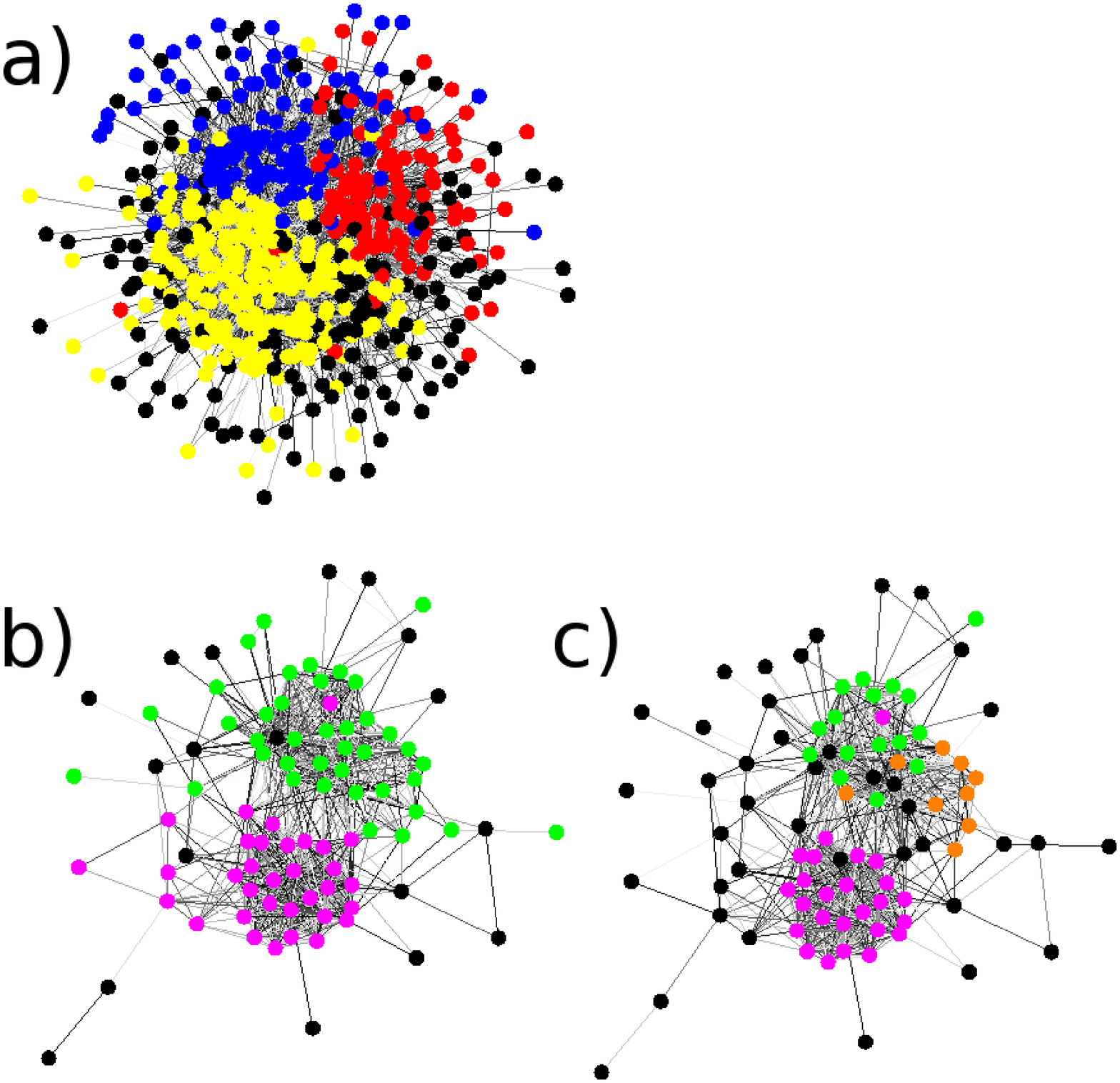}
\caption{\textbf{Examples of communities found.} a) A representation of a community in the $A$ network at resolution parameter value $\log(\lambda)=0$, with nodes (proteins) coloured according to the partition of this community at $\log(\lambda)=0.5$. The colours are the same as for Figure $1$a, where this group of proteins has labels roughly in the range $0 - 500$. Almost all of the nodes have some relationship to the ribosome. The proteins in the yellow community are mostly ribosomal subunits, those in the red community are mostly pre-cursors to and processors of the small ribosomal subunit, and those in the blue community have similar roles to those in the red community but for the large subunit. The shading of the links has no significance; its purpose is to ease visualisation. Black nodes are not located in one of the three largest communities discussed in the text. b) A representation of a community at $\log(\lambda)=0.5$, with nodes (proteins) coloured according to the partition of this community at $\log(\lambda)=0.75$. The proteins identified at the lower resolution almost all play some role in transcription initiation. At the higher resolution, more structure is revealed: the pink community consists mostly of proteins from the RNA polymerase II mediator complex and the green community mostly consists of proteins from the TFIID and SAGA complexes. c) The partition at a higher resolution ($\log(\lambda)=1.6$). The green community from b) has split into the SAGA complex (green) and the TFIID complex (orange). The names and descriptions of the proteins in these example communities are given in Additional File $2$. The node positions for visualisation were computed in the same way as for Figure $1$.}
\end{figure}

Consider the community at $\log(\lambda) = 0$ that is marked as the blue block in Figure $1$ for the $A$ network (over node labels approximately $0$ to $500$). This contains $528$ proteins and consists largely of proteins with some relationship to the ribosome (based on short protein descriptions found on the SGD website). Figure $4$a shows this community, where we have coloured nodes according to the community partition at the later partition $\log(\lambda) = 0.5$. The colours -- red, yellow, and blue -- are the same as in Figure $1$, where most of the community present at $\log(\lambda)=0$ has split into three communities at $\log(\lambda)=0.5$. The blue community consists of $107$ proteins, which are largely precursors to and processors of the large ribosomal unit. The red community consists of $95$ proteins, which have a similar function but for the small ribosomal subunit. The yellow community has $190$ proteins, $93$ of which are constituents of the ribosome and the remainder of which are either of unknown function or associate to the ribosome. We give short descriptions of the proteins in these communities in Additional File $2$. 

\mbox{}

An illustration of the biological relevance of community structure at three partitions is given in Figures $4$b and c. We show a community of $90$ proteins at $\log(\lambda)=0.5$, and display its partition into communities at b) $\log(\lambda)=0.75$ and c) $\log(\lambda)=1.6$. Almost all of the proteins in the community at $\log(\lambda)=0.5$ play some role in transcription initiation. At $\log(\lambda)=0.75$ this community has split into two main smaller communities: the pink community contains constituent proteins of the RNA polymerase II mediator complex and the green community contains components of the closely related SAGA and TFIID complexes. At $\log(\lambda)=1.6$, this second community has split into the SAGA and TFIID complexes.

\mbox{}

Multi-resolution community detection and characterisation is relevant both from the global viewpoint, where one can investigate the aggregate functional organisation of the proteome, and from the local perspective, where the community membership of particular proteins can be traced through changing resolution parameter. We thus now consider a protein-centred view of multi-resolution community detection. We consider, for an example protein, the properties of the communities to which it is assigned through changing resolution parameter, see Figure $5$. The size of the communities, their mean similarity under the $G$ and $C$ measures, and the mean clustering coefficient are shown. The protein is a member of the ESCRT-I complex. (Figure S$6$ in Additional File $1$ gives a further four examples.)  Note the very robust properties of the communities in the $A$ network over resolution parameter values of approximately $1 \leq \log(\lambda) \leq 2.5$, despite the tendency for them to be partitioned as $\lambda$ increases. At these resolutions, the protein is in the same community as other members of the complex, as well as a few other very closely associated proteins. Beyond $\log(\lambda) = 2.5$, the complex is broken up, as reflected in the drop in mean similarity values. The community present over $0.7 \leq \log(\lambda) \leq 1.4$ in the $P$ network contains many proteins associated to the complex (in addition to the complex itself). Above the step observable at $\log(\lambda) = 1.4$, only members of the complex are present. In Additional File $2$, we give the names and brief functional descriptions of proteins that occur in some of the same communities for this example, and the four other examples given in Additional File $1$. These five examples all show the following behaviour.
\begin{itemize}
\item In general, as would be expected, the size of the community to which a protein is assigned decreases with increasing resolution. There is often a large range of resolutions over which the community has constant size (which we have observed in practice to entail the same community across multiple resolutions). Such communities are particularly resilient to being split up at increasing resolutions, despite the tendency for them to be partitioned.
\item The community similarity under the $G$, $C$ and $M$ measures often shows a close correlation.
\item At higher resolutions, there tends to be a higher community similarity, as might be expected of a hierarchically organised system. This is, however, not always the case: community similarity can decrease at higher resolutions. In these instances, a group of proteins has been partitioned beyond the point at which function is shared, possibly through the exclusion of proteins involved in the same processes that do not necessarily directly interact with each other.
\item There is often a large overlap between the community membership in the $A$ and $P$ networks, but it can also be quite different. For example, in Additional File $1$ Figure S$6$c, the protein occurs with other proteins in the same complex in the $A$ network, whereas in the $P$ network it occurs with non-complex members which are nonetheless involved in the same process. The functional homogeneity of communities can also be different: sometimes the protein occurs in many functionally homogeneous community in the $A$ network and not the $P$, and sometimes vice versa. This is unsurprising given the very different nature of $A$ and $P$ interactions. By treating them separately, we are able to pick out both types of pattern. 
\end{itemize} 

\begin{figure}
\includegraphics[width=120mm]{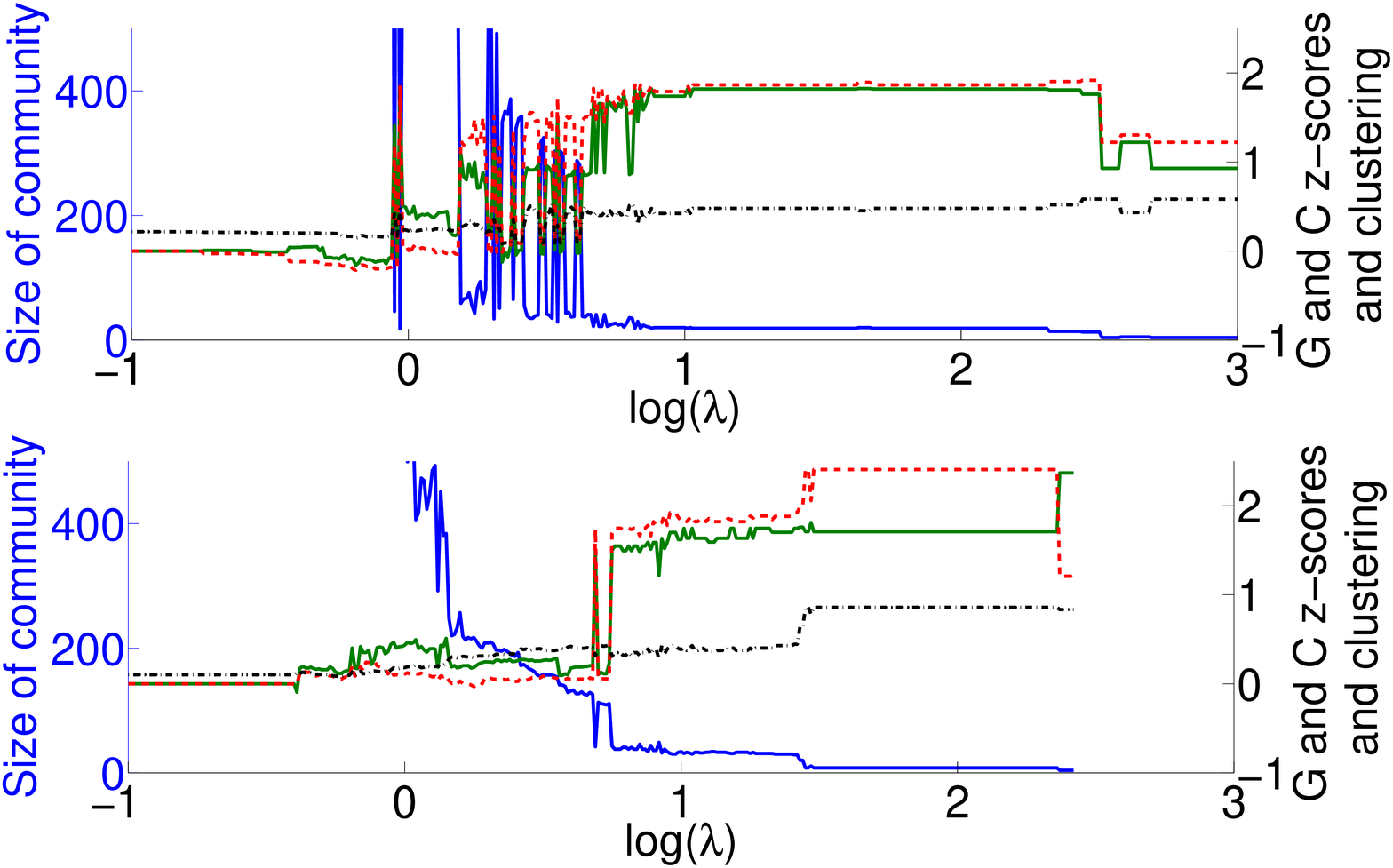}
\caption{\textbf{Tracing the community membership of a particular protein through changing resolution.} For the example protein YCL008C, we show the size (solid blue curve), mean clustering coefficient (dot-dashed black curve), mean $z$-score under the GO measure (solid green curve), and correlated growth measure (dashed red curve) with changing resolution for the $A$ network (top) and $P$ network (bottom). Long plateaus in these properties represent robust communities. We give further examples in Additional File $1$ Figures S$6$.}
\end{figure}

\subsection*{Use of topological properties to select functionally homogeneous communities}
\label{sec:topprops}
Almost all proteins are in functionally homogeneous communities at some value of the resolution parameter, and we therefore devise a method to swiftly identify these resolutions, especially if there is a dearth of functional information. We investigate whether any easily-calculated topological properties of the communities can act as indicators of functional homogeneity. Given a protein of interest we can then use such measures to quickly identify `good’ resolutions, without the need to assess functional homogeneity.

\mbox{}

We tested $26$ topological properties for their ability to predict functional homogeneity using the AUC metric (see Methods), and show our results in Table $4$. In general, the AUCs for the $P$ network are lower than those for the $A$ network, perhaps because there is more potentially usable information in the $A$ network as it is significantly denser (see Table $1$). 

\begin{table}
\begin{tabular}{|l||c|c|c||c|c|c|}
\hline
&\multicolumn{3}{|c||}{\textbf{$A$}}&\multicolumn{3}{c|}{\textbf{$P$}}\\ 	
Network topology measure & $G$ & $C$ & $M$ & $G$ & $C$ & $M$ \\
\hline
Mean degree   &  0.6476  &  0.6476  &  0.6142  &  0.5130  &  0.5373  &  0.5387 \\
Degree assortativity coefficient \cite{newman2002assortative}  &  0.6913  &  0.6913  &  0.6277  &  0.4799  &  0.5517  &  0.5181\\
\textbf{Clustering coefficient \cite{ldf2007characterization}}  &  \textbf{0.7186}  &  \textbf{0.7186}  &  \textbf{0.6613}  &  \textbf{0.5521}  &  \textbf{0.5829}  &  \textbf{0.5725}\\
Global mean Soffer clustering coefficient \cite{soffer2005network}   &  0.4857  &  0.4857  &  0.4819  &  0.3915  &  0.4735  &  0.4461\\
Local mean Soffer clustering coefficient \cite{soffer2005network}   &  0.4784  &  0.4784  &  0.4662  &  0.3892  &  0.4654  &  0.4540\\
Mean geodesic node betweenness centrality \cite{wasserman1994social}  &  0.4600  &  0.4600  &  0.4973  &  0.5045  &  0.5094  &  0.4959\\
Mean closeness centrality \cite{wasserman1994social}   &  0.5275  &  0.5275  &  0.5524  &  0.4877  &  0.4919  &  0.4815\\
Mean eigenvector centrality \cite{wasserman1994social}  &  0.5601  &  0.5601  &  0.5722  &  0.5312  &  0.5551  &  0.5246\\
Mean information centrality \cite{wasserman1994social} &  0.5191  &  0.5191  &  0.5429  &  0.5253  &  0.5456  &  0.5170\\
Mean geodesic distance  \cite{ldf2007characterization} &  0.3839  &  0.3839  &  0.3717  &  0.4274  &  0.4945  &  0.5066\\
Diameter \cite{wasserman1994social} &  0.4457  &  0.4457  &  0.4042  &  0.4366  &  0.5004  &  0.5079\\
Mean harmonic geodesic distance \cite{ldf2007characterization}  &  0.4088  &  0.4088  &  0.4042  &  0.5024  &  0.4834  &  0.4995\\
Energy \cite{ldf2007characterization}  &  0.5237  &  0.5237  &  0.4982  &  0.4568  &  0.4976  &  0.5114\\
Entropy \cite{ldf2007characterization}  &  0.5655  &  0.5655  &  0.5327  &  0.5077  &  0.5127  &  0.5280\\
Off-diagonal complexity \cite{kim2008complex}  &  0.5941  &  0.5941  &  0.5457  &  0.5081  &  0.5054  &  0.5237\\
Cyclomatic number \cite{kim2008complex}  &  0.6331  &  0.6331  &  0.5733  &  0.5173  &  0.5300  &  0.5425\\
Connectivity \cite{kim2008complex}   &  0.6437  &  0.6437  &  0.5766  &  0.5245  &  0.5334  &  0.5468\\
Number of spanning trees \cite{kim2008complex}  &  0.4525  &  0.4525  &  0.4531  &  0.4451  &  0.4516  &  0.4491\\
Medium articulation \cite{kim2008complex}  &  0.5659  &  0.5659  &  0.4463  &  0.5295  &  0.5070  &  0.5592\\
Efficiency complexity \cite{kim2008complex}  &  0.5316  &  0.5316  &  0.5343  &  0.4911  &  0.4945  &  0.4982\\
Graph index complexity \cite{kim2008complex}  &  0.6564  &  0.6564  &  0.6492  &  0.5211  &  0.5469  &  0.5250\\
Density  &  0.6541  &  0.6541  &  0.6553  &  0.5277  &  0.5676  &  0.5235\\
Efficiency \cite{latora2001efficient}  &  0.5790  &  0.5790  &  0.5896  &  0.4964  &  0.5071  &  0.4865\\
Fraction of articulation vertices \cite{tsukiyama1980algorithm}  &  0.5065  &  0.5065  &  0.5028  &  0.5216  &  0.5062  &  0.5091\\
Largest eigenvalue  &  0.6054  &  0.6054  &  0.5663  &  0.4941  &  0.5041  &  0.5185\\
Rich club coefficient \cite{colizza2006detecting}  &  0.5428  &  0.5428  &  0.5896  &  0.4988  &  0.5209  &  0.4868\\
\hline
\end{tabular}
\caption{\textbf{Topological metrics tested and AUCs.}
The network topology measures tested and their associated AUCs. We report the results for using each of these as a predictor for functional homogeneity as judged under the three measures of functional similarity (GO, $G$, correlated growth rates, $C$, and MIPS, $M$) for both the $A$ and $P$ networks. The AUCs are given as the average performance over the range $0 \leq \log(\lambda) \leq 3$. The clustering coefficient (definition given in the text, equation \ref{eq:clustering}) is the best predictor in all cases. (The topological properties were computed from code developed by Gabriel Villar.)}
\end{table}

\mbox{}

We find that clustering coefficient is the most useful of the topological properties tested in the prediction of functional homogeneity for all three similarity measures and in both the $A$ and $P$ networks. The clustering coefficient of a network is a measure of the mean local clustering around nodes: A node has a high clustering coefficient, $c$, if its neighbours are also neighbours of each other \cite{clustering, siam03}. It is defined for each node as
\begin{equation}
c = \frac{3N_{{\rm triangle}}}{N_{{\rm triple}}}, 
\label{eq:clustering}
\end{equation}
where $N_{{\rm triangle}}$ is the number of triangles of which the node is a member, and $N_{{\rm triple}}$ is the number of connected triples of which the node is a member. (A connected triple is a single node with edges running to an unordered pair of other nodes.) Figure $6$ shows the ROC curve for using the mean clustering coefficient of nodes in a community as a predictor of functional homogeneity for each of the three similarity measures in the $A$ network. (See Methods for a description of the construction. The corresponding Figure for the $P$ network is given in Additional File $1$, Figure S$7$.)

\begin{figure}
\includegraphics[width=120mm]{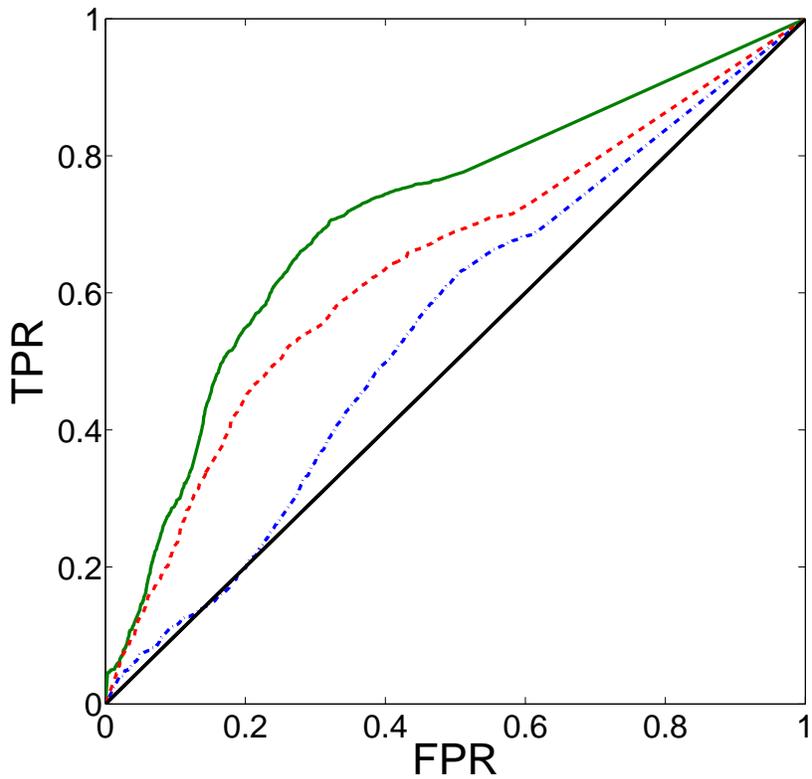}
\caption{\textbf{ROC curve for using mean clustering coefficient to pick out functionally homogeneous communities in the $A$ network.} The Receiver Operating Characteristic (ROC) curve for using mean clustering coefficient as a predictor of functional homogeneity under the GO measure (solid green curve), MIPS measure (dot-dashed blue curve) and correlated growth measure (dashed red curve). We plot the false positive rate (FPR) versus the true positive rate (TPR). A random classifier would give the solid black line. For the GO measure, a true positive rate of $70\%$ is achievable with a false positive rate of $30\%$. The best predictive ability is achieved for the GO measure, and the worst for the MIPS measure (see Table $4$ for areas under the curves (AUCs)).}
\end{figure}

\mbox{}

There is some element of discretion for annotating $A$ type interactions, i.e. deciding which pairs to list interactions between following experiments, with the principle competing models referred to as `matrix' and `spoke' \cite{bader02}. This choice could cause artefactual topological features, so the extent to which we find particular topological features correlating with functional homogeneity could be sensitive to annotation choice. We are therefore encouraged that the same trends in predictive ability are evident in the $P$ network, for which there is no such element of discretion.

\mbox{}

As can be seen from Figure $5$ and the figures in Additional File $1$ Figure S$6$, clustering appears to be a good proxy for functional homogeneity when looking at individual proteins, and in the absence of much functional information could guide which resolution(s) should be targeted for investigation.

\section*{Conclusions}

If protein interaction networks are to aid understanding of how biological function emerges from the concerted action of many proteins, then it is crucial to explore connections between network structure and biological function. In this paper we investigate how the function of sets of proteins varies with network community structure of yeast at multiple resolutions.

\mbox{}

We find that community structure does indeed help identify sets of proteins that act together, and that this connection between network structure and biological function depends on what network scales are probed. We do not expect there to be any single scale of interest in this middle-scale structure of the protein interaction network; although previous studies have applied community detection algorithms to protein interaction networks, no study to our knowledge has investigated this structure at multiple resolutions. We find that $4652$ of $4980$ proteins in the $A$ network, and $5647$ of $5669$ proteins in the $P$ network, are in functionally homogeneous communities at some value of the resolution parameter as judged under the GO similarity measure. The number of proteins in functionally homogeneous communities peaks at about $\lambda = 3$ for the $A$ network (which is beyond the standard `modularity' resolution of $\lambda = 1$). For the $P$ network the peak is less pronounced, with the actual maximum occurring at $\lambda = 7$ (i.e. $\log(\lambda) = 0.86$). These findings emphasise that there are different scales of interest in the community structure of protein interaction networks, and that the one of primary interest will depend on which proteins and processes one is investigating. For some protein types, there are natural resolutions, at which more proteins of that type are assigned to functionally homogeneous communities. We also find that proteins involved in some processes are much more likely to be in functionally homogeneous communities than others. For example we find for both networks and across a range of resolutions that approximately $70-80\%$ of proteins involved in chromosome organisation compared to $40\%$ involved in lipid metabolism are in functionally homogeneous communities.

\mbox{}

Having a good measure of functional homogeneity is central for our analysis. We approach this issue by using three different characterisations of functional similarity: two based on the GO and MIPS structured vocabularies respectively and one based on the growth rates of gene knock-out strains under different chemical conditions \cite{hillenmeyer2008chemical} (an independent and objective characterization of biological function). The prevalent method in the literature for assessing functional homogeneity of a group of proteins is inappropriate for communities, as the number of interacting pairs in a group must be taken into consideration. By defining similarity at the pairwise level, we have developed a fair test of functional homogeneity through a comparison of interacting pairs. We also capture the aggregate functional similarity of two proteins, overcoming the need to assess functional homogeneity on a term by term basis (although this is, of course, also possible once communities of particular interest have been identified). Our tests of functional homogeneity (which are not statistical tests in the conventional sense because of our desire to exclude the effects of sample size) using the three measures of similarity show a high level of agreement with each other, giving us confidence in our chosen measures of functional similarity. 

\mbox{}

Throughout this study, we have investigated two separate yeast protein interaction networks: that based on associations (the $A$ network; mostly TAP-like data), and that based on physical associations (the $P$ network; mostly yeast-two-hybrid data). We find that the two networks have similar properties with respect to their community structure, despite their very different global topological properties. Rather than regarding the yeast-two-hybrid data as of an inferior quality \cite{y2h_worse}, we start from the basis that it is of a fundamentally different type and should thus be treated separately. We find similar percentages of functionally homogeneous communities in both networks. 

\mbox{}

As we have found a connection between network communities and biological function, we can use observed community structure to predict aspects of biological function. We find in particular that communities with a high mean clustering coefficient are far more likely to be functionally homogeneous than those with a lower one. The mean clustering coefficient of nodes within a community can therefore be used to predict that a group of proteins is functionally homogeneous, even in cases where our current knowledge does not allow us to infer this on the basis of functional annotations alone. These results give insights into the relationships between the structural and functional organisation of the cell considering the whole proteome. 

\mbox{}

We have also illustrated the utility of our framework for biologists who are interested in a particular protein. In a chosen interaction network, one can determine the community membership of the protein of interest at multiple resolutions. Even in the dearth of functional information, the easily-calculated clustering coefficient can be computed to suggest resolutions of particular interest. 

\mbox{}

In conclusion, we have linked the community structure of a protein interaction network with biological function by probing different scales of network structure. The identified communities are candidates for biological modules within the cell. We have also illustrated how this connection can be used to select groups of proteins that likely participate in similar biological functions.

\section*{Authors contributions}
    All four authors conceived of the study, and ACFL carried it out.

\section*{Acknowledgements}
  \ifthenelse{\boolean{publ}}{\small}{}
We thank Gesine Reinert, Simon Myers, Sumeet Agarwal, Dan Fenn, and Peter Mucha for useful discussions. We thank Gabriel Villar for implementing the network measures, and Amanda Traud for implementation of the Kamada-Kawai visualisation code (which we modified for use here), which can be found at \url{http://netwiki.amath.unc.edu/VisComms/VisComms}.


{\ifthenelse{\boolean{publ}}{\footnotesize}{\small}

\begin{thebibliography}{10}
\providecommand{\url}[1]{[#1]}
\providecommand{\urlprefix}{}

\bibitem{shoemaker07a}
Shoemaker BA, Panchenko AR: \textbf{Deciphering protein--protein interactions
  Part I Experimental techniques and databases}. \emph{PLoS Computational
  Biology} 2007, \textbf{3}(3):337--334.

\bibitem{Tarassov08}
Tarassov K, Messier V, Landry CR, Radinovic S, Molina MM, Shames I, Malitskaya
  Y, Vogel J, Bussey H, Michnick SW: \textbf{An in vivo map of the yeast
  protein interactome}. \emph{Science} 2008, \textbf{320}(5882):1465--1470.

\bibitem{yu08}
Yu H, Braun P, Yildirim MA, Lemmens I, Venkatesan K, Sahalie J,
  Hirozane-Kishikawa T, Gebreab F, Li N, Simonis N, Hao T, Rual JF, Dricot A,
  Vazquez A, Murray RR, Simon C, Tardivo L, Tam S, Svrzikapa N, Fan C, de~Smet
  AS, Motyl A, Hudson ME, Park J, Xin X, Cusick ME, Moore T, Boone C, Snyder M,
  Roth FP, Barab{\'a}si A-L, Tavernier J, Hill DE, Vidal M: \textbf{High-quality
  binary protein interaction map of the yeast interactome network}.
  \emph{Science} 2008, \textbf{322}(5898):104--110.

\bibitem{Hartwell}
Hartwell LH, Hopfield JJ, Leibler S, Murray AW: \textbf{{From molecular to
  modular cell biology}}. \emph{Nature} 1999, \textbf{402}(6761):C4--C52.

\bibitem{Ravasz}
Ravasz E, Somera AL, Mongru DA, Oltvai ZN, Barab{\'a}si A-L:
  \textbf{{Hierarchical organization of modularity in metabolic networks}}.
  \emph{Science} 2002, \textbf{297}(5586):1551--1555.

\bibitem{han2004edo}
Han JDJ, Bertin N, Hao T, Goldberg DS, Berriz GF, Zhang LV, Dupuy D, Walhout
  AJM, Cusick ME, Roth FP, et~al.: \textbf{{Evidence for dynamically organized
  modularity in the yeast protein-protein interaction network}}. \emph{Nature}
  2004, \textbf{430}(6995):88--93.

\bibitem{alon2007isb}
Alon U: \emph{{An Introduction to Systems Biology: Design Principles of
  Biological Circuits}}. Chapman \& Hall/CRC 2007.

\bibitem{yook2004functional}
Yook SH, Oltvai ZN, Barab{\'a}si A-L: \textbf{{Functional and topological
  characterization of protein interaction networks}}. \emph{Proteomics} 2004,
  \textbf{4}(4):928--942.

\bibitem{hierarchicalmodularity1}
Rives AW, Galitski T: \textbf{{Modular organization of cellular networks}}.
  \emph{Proceedings of the National Academy of Sciences} 2003,
  \textbf{100}(3):1128--1133.

\bibitem{hierarchicalmodularity2}
Bachman P, Liu Y: \textbf{{Structure discovery in PPI networks using
  pattern-based network decomposition}}. \emph{Bioinformatics} 2009,
  \textbf{25}(14):1814--1821.

\bibitem{comnotices}
Porter MA, Onnela J-P, Mucha PJ: \textbf{{Communities in networks}}.
  \emph{Notices of the American Mathematical Society} 2009,
  \textbf{56}(9):1082--1097, 1164--1166.

\bibitem{santo_review}
Fortunato S: \textbf{Community detection in graphs}. \emph{Physics Reports}
  2010, \textbf{486}:75--174.

\bibitem{bu2003tsa}
Bu D, Zhao Y, Cai L, Xue H, Zhu X, Lu H, Zhang J, Sun S, Ling L, Zhang N,
  et~al.: \textbf{{Topological structure analysis of the protein-protein
  interaction network in budding yeast }}. \emph{Nucleic Acids Research} 2003,
  \textbf{31}(9):2443--2450.

\bibitem{pereiraleal2004dfm}
Pereira-Leal JB, Enright AJ, Ouzounis CA: \textbf{{Detection of functional
  modules from protein interaction networks}}. \emph{Proteins: Structure,
  Function and Genetics} 2004, \textbf{54}:49--57.

\bibitem{GOenrich}
Dunn R, Dudbridge F, Sanderson CM: \textbf{{The use of edge-betweenness
  clustering to investigate biological function in protein interaction
  networks}}. \emph{BMC Bioinformatics} 2005, \textbf{6}:39.

\bibitem{chen2006dfm}
Chen J, Yuan B: \textbf{{Detecting functional modules in the yeast
  protein-protein interaction network}}. \emph{Bioinformatics} 2006,
  \textbf{22}(18):2283--2290.

\bibitem{luo2007mop}
Luo F, Yang Y, Chen CF, Chang R, Zhou J, Scheuermann RH: \textbf{{Modular
  organization of protein interaction networks}}. \emph{Bioinformatics} 2007,
  \textbf{23}(2):207--214.

\bibitem{mete2008saf}
Mete M, Tang F, Xu X, Yuruk N: \textbf{{A structural approach for finding
  functional modules from large biological networks}}. \emph{BMC
  Bioinformatics} 2008, \textbf{9}:S19.

\bibitem{li2008gtm}
Li M, Wang J, Chen J: \textbf{{A graph-theoretic method for mining overlapping
  functional modules in protein interaction networks}}. \emph{Lecture Notes in
  Bioinformatics} 2008, \textbf{4983}:208--219.

\bibitem{GO}
Ashburner M, Ball CA, Blake JA, Botstein D, Butler H, Cherry JM, Davis AP,
  Dolinski K, Dwight SS, Eppig JT, et~al.: \textbf{{Gene Ontology: Tool for the
  unification of biology}}. \emph{Nature Genetics} 2000, \textbf{25}:25--29.

\bibitem{MIPS}
Mewes HW, Frishman D, Guldener U, Mannhaupt G, Mayer K, Mokrejs M, Morgenstern
  B, Munsterkotter M, Rudd S, Weil B: \textbf{{MIPS: A database for genomes and
  protein sequences }}. \emph{Nucleic Acids Research} 2002, \textbf{30}:31--34.

\bibitem{reslimit}
Fortunato S, Barthelemy M: \textbf{{Resolution limit in community detection}}.
  \emph{Proceedings of the National Academy of Sciences} 2007,
  \textbf{104}:36--41.

\bibitem{RandB}
Reichardt J, Bornholdt S: \textbf{{Statistical mechanics of community
  detection}}. \emph{Physical Review E} 2006, \textbf{74}:16110.

\bibitem{kumpula2007lra}
Kumpula JM, Saram{\"a}ki J, Kaski K, Kert{\'e}sz J: \textbf{{Limited resolution
  and multiresolution methods in complex network community detection}}.
  \emph{Fluctuation and Noise Letters} 2007, \textbf{7}(3):L209--L214.

\bibitem{heimo2008dmd}
Heimo T, Kumpula J, Kaski K, Saramaki J: \textbf{{Detecting modules in dense
  weighted networks with the Potts method}}. \emph{Journal of Statistical
  Mechanics: Theory and Experiment} 2008, (P08007).

\bibitem{arenas2008asc}
Arenas A, Fern{\'a}ndez A, G{\'o}mez S: \textbf{{Analysis of the structure of
  complex networks at different resolution levels}}. \emph{New Journal of
  Physics} 2008, \textbf{10}:053039.

\bibitem{blondel2008fuc}
Blondel VD, Guillaume JL, Lambiotte R: \textbf{{Fast unfolding of communities
  in large networks}}. \emph{Journal of Statistical Mechanics: Theory and
  Experiment} 2008, (P10008).

\bibitem{pu2007identifying}
Pu S, Vlasblom J, Emili A, Greenblatt J, Wodak SJ: \textbf{{Identifying
  functional modules in the physical interactome of Saccharomyces cerevisiae}}.
  \emph{Proteomics} 2007, \textbf{7}(6):944--960.

\bibitem{y2h_worse}
Song J, Singh M: \textbf{{How and when should interactome-derived clusters be
  used to predict functional modules and protein function?}}
  \emph{Bioinformatics} 2009, \textbf{25}(23):3143--3150.

\bibitem{hillenmeyer2008chemical}
Hillenmeyer ME, Fung E, Wildenhain J, Pierce SE, Hoon S, Lee W, Proctor M,
  St~Onge RP, Tyers M, Koller D, et~al.: \textbf{{The chemical genomic portrait
  of yeast: uncovering a phenotype for all genes}}. \emph{Science} 2008,
  \textbf{320}(5874):362.

\bibitem{MI}
Hermjakob H, Montecchi-Palazzi L, Bader G, Wojcik J, Salwinski L, Ceol A, Moore
  S, Orchard S, Sarkans U, von Mering C, et~al.: \textbf{{The HUPO PSI's
  molecular interaction format—a community standard for the representation of
  protein interaction data}}. \emph{Nature Biotechnology} 2004,
  \textbf{22}(2):177--183.

\bibitem{li04}
Li S, Armstrong CM, Bertin N, Ge H, Milstein S, Boxem M, Vidalain PO, Han JD,
  Chesneau A, Hao T, Goldberg DS, Li N, Martinez M, Rual JF, Lamesch P, Xu L,
  Tewari M, Wong SL, Zhang LV, Berriz GF, Jacotot L, Vaglio P, Reboul J,
  Hirozane-Kishikawa T, Li Q, Gabel HW, Elewa A, Baumgartner B, Rose DJ, Yu H,
  Bosak S, Sequerra R, Fraser A, Mango SE, Saxton WM, Strome S, Van Den~Heuvel
  S, Piano F, Vandenhaute J, Sardet C, Gerstein M, Doucette-Stamm L, Gunsalus
  KC, Harper JW, Cusick ME, Roth FP, Hill DE, Vidal M: \textbf{A map of the
  interactome network of the metazoan C elegans}. \emph{Science} 2004,
  \textbf{303}(5657):540--543.

\bibitem{collins08}
Collins MO, Choudhary JS: \textbf{Mapping multiprotein complexes by affinity
  purification and mass spectrometry}. \emph{Current Opinion in Biotechnology}
  2008, \textbf{19}(4):324--330, \urlprefix\url{[http://www hubmed org/display
  cgi?uids=18598764]}.

\bibitem{vonmering02}
von Mering C, Krause R, Snel B, Cornell M, Oliver SG, Fields S, Bork P:
  \textbf{Comparative assessment of large-scale data sets of protein-protein
  interactions}. \emph{Nature} 2002, \textbf{417}(6887):399--403.

\bibitem{stark2006biogrid}
Stark C, Breitkreutz BJ, Reguly T, Boucher L, Breitkreutz A, Tyers M:
  \textbf{{BioGRID: a general repository for interaction datasets}}.
  \emph{Nucleic acids research} 2006, \textbf{34}(Database Issue):D535.

\bibitem{IntAct}
Kerrien S, Alam-Faruque Y, Aranda B, Bancarz I, Bridge A, Derow C, Dimmer E,
  Feuermann M, Friedrichsen A, Huntley R, et~al.: \textbf{{IntAct--open source
  resource for molecular interaction data}}. \emph{Nucleic acids research}
  2007, \textbf{35}(Database issue):D561.

\bibitem{zanzoni02}
Zanzoni A, Montecchi-Palazzi L, Quondam G, Helmer-Citterich M, Cesareni G:
  \textbf{MINT: a Molecular INTeraction database}. \emph{FEBS Letters} 2002,
  \textbf{513}:135--140.

\bibitem{NP}
Hastings MB: \textbf{{Community detection as an inference problem}}.
  \emph{Physical Review E} 2006, \textbf{74}(3):35102.

\bibitem{brandes2006mmh}
Brandes U, Delling D, Gaertler M, Goerke R, Hoefer M, Nikoloski Z, Wagner D:
  \textbf{On modularity clustering}. \emph{IEEE Transactions on Knowledge and
  Data Engineering} 2008, \textbf{20}(2):172--188.

\bibitem{santo09}
Lancichinetti A, Fortunato S: \textbf{Community detection algorithms: a
  comparative analysis}. \emph{Physical Review E} 2009, \textbf{80}:056117.

\bibitem{newmanpre}
Newman MEJ: \textbf{{Finding community structure in networks using the
  eigenvectors of matrices}}. \emph{Physical Review E} 2006,
  \textbf{74}(3):36104.

\bibitem{cherry1998sgd}
Cherry JM, Adler C, Ball C, Chervitz SA, Dwight SS, Hester ET, Jia Y, Juvik G,
  Roe T, Schroeder M, et~al.: \textbf{{SGD: Saccharomyces genome database}}.
  \emph{Nucleic Acids Research} 1998, \textbf{26}:73.

\bibitem{pandey}
Pandey J, Koyuturk M, Subramaniam S, et~al.: \textbf{Functional coherence in
  domain interaction networks}. \emph{Bioinformatics} 2008,
  \textbf{24}:I28--I34.

\bibitem{boyle2004go}
Boyle EI, Weng S, Gollub J, Jin H, Botstein D, Cherry JM, Sherlock G:
  \textbf{{GO:: TermFinder--open source software for accessing Gene Ontology
  information and finding significantly enriched Gene Ontology terms associated
  with a list of genes}}. \emph{Bioinformatics} 2004, \textbf{20}(18):3710.

\bibitem{stats_book}
Mendenhall W, Beaver RJ, Beaver BM: \emph{{Introduction to Probability and
  Statistics}}. Brooks/Cole 2008.

\bibitem{fawcett06}
Fawcett T: \textbf{An introduction to ROC analysis}. \emph{Pattern Recognition
  Letters} 2006, \textbf{27}(8):861--874.

\bibitem{morethanmodules}
{Pinkert, S and Schultz, J and Reichardt, J}: \textbf{Protein Interaction
  Networks - More than mere modules}. \emph{PLoS Computational Biology} 2010,
  \textbf{6}:e1000659.

\bibitem{GO_slim}
Hong EL, Balakrishnan R, Dong Q, Christie KR, Park J, Binkley G, Costanzo MC,
  Dwight SS, Engel SR, Fisk DG, et~al.: \textbf{{Gene Ontology annotations at
  SGD: new data sources and annotation methods}}. \emph{Nucleic acids research}
  2008, \textbf{36}(Database issue):D577.

\bibitem{clustering}
Watts DJ, Strogatz SH: \textbf{{Collective dynamics of
  ‘small-world’networks}}. \emph{Nature} 1998, \textbf{393}(6684):440--442.

\bibitem{siam03}
Newman MEJ: \textbf{The structure and function of complex networks}. \emph{SIAM
  Review} 2003, \textbf{45}:167--256.

\bibitem{bader02}
Bader GD, Hogue CW: \textbf{Analyzing yeast protein-protein interaction data
  obtained from different sources}. \emph{Nature Biotechnology} 2002,
  \textbf{20}(10):991--997.

\bibitem{kamada}
Kamada T, Kawai S: \textbf{{An algorithm for drawing general undirected
  graphs}}. \emph{Information processing letters} 1989, \textbf{31}:7--15.

\bibitem{newman2002assortative}
Newman MEJ: \textbf{{Assortative mixing in networks}}. \emph{Physical Review
  Letters} 2002, \textbf{89}(20):208701.

\bibitem{ldf2007characterization}
Costa LD, Rodrigues FA, Travieso G, Boas PRV: \textbf{{Characterization of
  complex networks: A survey of measurements}}. \emph{Advances in Physics}
  2007, \textbf{56}:167--242.

\bibitem{soffer2005network}
Soffer SN, V{\'a}zquez A: \textbf{{Network clustering coefficient without
  degree-correlation biases}}. \emph{Physical Review E} 2005,
  \textbf{71}(5):57101.

\bibitem{wasserman1994social}
Wasserman S, Faust K: \emph{{Social Network Analysis: Methods and
  Applications}}. Cambridge University Press 1994.

\bibitem{kim2008complex}
Kim J, Wilhelm T: \textbf{{What is a complex graph?}} \emph{Physica A:
  Statistical Mechanics and its Applications} 2008, \textbf{387}:2637--2652.

\bibitem{latora2001efficient}
Latora V, Marchiori M: \textbf{{Efficient behavior of small-world networks}}.
  \emph{Physical Review Letters} 2001, \textbf{87}(19):198701.

\bibitem{tsukiyama1980algorithm}
Tsukiyama S, Shirakawa I, Ozaki H, Ariyoshi H: \textbf{{An algorithm to
  enumerate all cutsets of a graph in linear time per cutset}}. \emph{Journal
  of the ACM} 1980, \textbf{27}(4):619--632.

\bibitem{colizza2006detecting}
Colizza V, Flammini A, Serrano MA, Vespignani A: \textbf{{Detecting rich-club
  ordering in complex networks}}. \emph{Nature Physics} 2006,
  \textbf{2}:110--115.

\end{thebibliography}

}


\ifthenelse{\boolean{publ}}{\end{multicols}}{}

\end{bmcformat}
\end{document}